  [2009/10/30 v0 ECT* Trento 2009 - article-version driver (PGR)]
\documentclass{ws-mpla}
\usepackage{Ratcliffe}
\title{NON-UNIVERSALITY AND EVOLUTION\\ OF THE SIVERS FUNCTION}
\author{\footnotesize PHILIP G. RATCLIFFE}
\address{%
  Dipartimento di Fisica e Matematica,
  Universit\`{a} degli Studi dell'Insubria, Como
  \\ \tstrut
  and
  \\
  Istituto Nazionale di Fisica Nucleare,
  Sezione di Milano--Bicocca
}
\author{\footnotesize OLEG TERYAEV}
\address{%
  Bogoliubov Laboratory of Theoretical Physics,
  Joint Institute for Nuclear Research, Dubna
}
\begin{document}
\markboth{P.G.~Ratcliffe and O.~Teryaev}
         {Non-Universality and Evolution of the Sivers Function}
\maketitle
\begin{abstract}
  We examine the large-$x$ QCD evolution of the twist-three gluonic-pole strength defining an effective T-odd Sivers function, where evolution of the T-even transverse-spin DIS structure function $g_2$ is multiplicative. The result corresponds to a colour-factor modified spin-averaged twist-two evolution.
\end{abstract}
\ccode{PACS Nos.: 13.88.+e}
\section{Introduction}

\subsection{Motivation}
Single-spin asymmetries (SSA's) have long been an enigma in high-energy hadronic physics. Before being first observed experimentally, hadronic SSA's were predicted to be small for a variety of reasons. Experimentally, however, SSA's turn out to be large in many hadronic processes. It was also long held that such asymmetries would eventually vanish at large energy and/or $p_T^{}$. Again, however, the SSA's so far observed show no signs of high-energy suppression.

\subsection{What is transverse spin?}
Transverse here indicates a spin vector perpendicular to particle momentum (in contrast to parallel as in, \emph{e.g.}, the case of the DIS structure function $g_1$). Note that the existence of transverse polarisation is itself independent of particle masses---\emph{cf.} the natural ($\sim9\,\%$) LEP beam polarisation. The problem of (small) masses only arises when seeking measurable transverse-spin effects, which require spin flip.

\subsection{Single-spin asymmetries}
Generically, SSA's reflect correlations of the form $\vec{s}\cdot(\vec{p}\vprod\vec{k})$, where $\vec{s}$ is some particle polarisation vector, while $\vec{p}$ and $\vec{k}$ are initial/final particle/jet momenta. A typical example might be: $\vec{p}$ the beam direction, $\vec{s}$ the target polarisation (transverse with respect to $\vec{p}$) and $\vec{k}$ the final-state particle direction (necessarily out of the $\vec{p}$--$\vec{s}$ plane). Polarisations involved in SSA's must therefore be transverse (although there are certain special exceptions). It is more convenient to use an helicity basis via the transformation
\begin{align}
  \ket{\uparrow/\downarrow} =
  \tfrac1{\sqrtno2\,} \left[ \strut\, \ket{+} \pm \im \ket{-} \right].
\end{align}
Any such asymmetry then takes on the (schematic) form
\begin{align}
  \mathcal{A}_N
  \sim
  \frac{\braket{\uparrow|\uparrow}-\braket{\downarrow|\downarrow}}
       {\braket{\uparrow|\uparrow}+\braket{\downarrow|\downarrow}}
  \sim
  \frac{2\Im\braket{+|-}}{\braket{+|+}+\braket{-|-}}
\end{align}
The appearance of both $\ket{+}$ and $\ket{-}$ in the numerator indicates the presence of a helicity-flip amplitude. The precise form of the numerator indicates interference between two helicity amplitudes: one helicity-flip and one non-flip, with a relative phase difference, the imaginary phase implying na\"{\i}ve T-odd processes.

\citet{Kane:1978nd} realised that in the massless (or high-energy) limit and the Born approximation a gauge theory such as QCD cannot furnish either requirement: massless fermion helicity is conserved and tree diagram amplitudes are always real. This led to the famous statement \cite{Kane:1978nd}: ``\dots\ \emph{observation of significant polarizations in the above reactions would contradict either QCD or its applicability.}''

As we now know, large asymmetries were found, but QCD nevertheless survived! \citet*{Efremov:1984ip} discovered one way out within the context of perturbative QCD. Consideration of the three-parton correlators involved in, \emph{e.g.}, $g_2$, leads to the following crucial observations: the relevant mass scale is not that of the current quark, but of the hadron; the pseudo-two-loop nature of the diagrams leads to an imaginary part in certain regions of partonic phase space~\cite{Efremov:1981sh}.

However, it took time before progress was made and the richness of the available structure was fully exploited---see \cite{Qiu:1991pp.x}. Indeed, it turns out that there are a variety of mechanisms to produce SSA's:
\begin{itemize}
\item
Transversity: hadron helicity flip is correlated to quark flip.
Chirality conservation requires another T-odd (distribution or fragmentation) function.
\item
Internal quark motion: correlation between the transverse polarisation of a quark and its own transverse momentum.
This corresponds to the \citeauthor{Sivers:1989cc} function \cite{Sivers:1989cc} and requires orbital angular momentum together with soft-gluon exchange.
\item
\mbox{Twist-3} transverse-spin dependent three-parton correlators (\emph{cf.} $g_2$):  the pseudo two-loop nature provides effective spin flip and an imaginary part via pole terms.
\end{itemize}
The second and third mechanisms turn out to be related.

\section{Single-Spin Asymmetries}

\subsection{Single-hadron production}
As a consequence of the multiplicity of underlying mechanisms, there are various types of distribution and fragmentation functions that can be active in generating SSA's:
\begin{itemize}
\item
higher-twist distribution and fragmentation functions,
\item
$k_T$-dependent distribution and fragmentation functions,
\item
interference fragmentation functions,
\item
higher-spin functions, \emph{e.g.}, vector-meson fragmentation functions.
\end{itemize}
Consider the case of single-hadron production with one initial-state transversely polarised hadron:
\begin{equation}
  A^\uparrow(P_A) + B(P_B) \to h(P_h) + X,
\end{equation}
where hadron $A$ is transversely polarised while $B$ is not. The unpolarised (or spinless) hadron $h$ is produced at large transverse momentum $\Vec{P}_{hT}$ and PQCD is thus applicable. Typically, $A$ and $B$ are protons while $h$ may be a pion or kaon \emph{etc}. The following SSA may be measured:
\begin{align}
  A_{T}^h =
  \frac{\D\sigma(\Vec{S}_T) - \D\sigma(-\Vec{S}_T)}
       {\D\sigma(\Vec{S}_T) + \D\sigma(-\Vec{S}_T)}
\end{align}

\begin{figure}[hbt]
  \centering
  \includegraphics[width=0.6\textwidth]{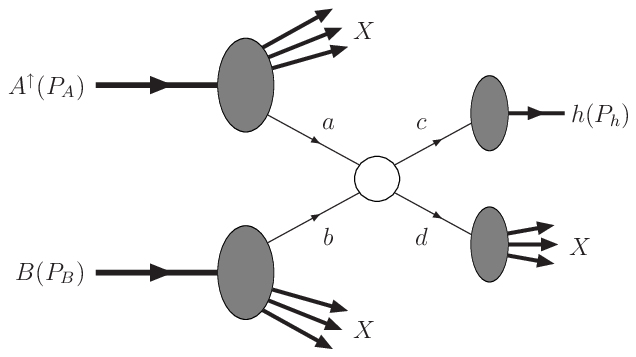}
  \caption{%
    Factorisation in single-hadron production with a transversely polarised hadron.
  }
  \label{fig:fact}
\end{figure}

\begin{figure}[hbt]
  \centering
  \raisebox{13mm}{%
    $\mathcal{M}_{\alpha\beta\gamma\delta}$ \quad = \quad
  }
  \includegraphics[width=0.3\textwidth]{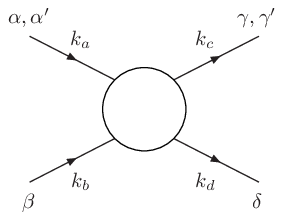}
  \caption{%
    The hard partonic amplitude, $\alpha\beta\gamma\delta$ are Dirac indices.
  }
  \label{fig:matel}
\end{figure}

Assuming standard factorisation to hold, the differential cross-section for such a process may be written formally as (\emph{cf.} Fig.~\ref{fig:fact})
\begin{align}
  \qquad
  \D\sigma =
  \sum_{abc} \sum_{\alpha\alpha'\gamma\gamma'}
  \rho^a_{\alpha'\alpha} \,
  f_a(x_a) \otimes
  f_b(x_b) \otimes
  \D\hat\sigma_{\alpha\alpha'\gamma\gamma'} \otimes
  \mathcal{D}_{h/c}^{\gamma'\gamma}(z)
  ,
\end{align}
where $f_a$ ($f_b$) is the density of parton type $a$ ($b$) inside hadron $A$ ($B$), $\rho^a_{\alpha\alpha'}$ is the spin density matrix for parton $a$, $\mathcal{D}_{h/c}^{\gamma\gamma'}$ is the fragmentation matrix for parton $c$ into the final hadron $h$ and $\D\hat\sigma_{\alpha\alpha'\gamma\gamma'}$ is the partonic cross-section:
\begin{align}
  \left(
    \frac{\D\hat\sigma}{\D\hat{t}}
  \right)_{\!\!\alpha\alpha'\gamma\gamma'}
  =
  \frac1{16\pi\hat{s}^2} \, \frac12 \, \sum_{\beta\delta}
  \mathcal{M}^{\vphantom*}_{\alpha\beta\gamma\delta} \,
  \mathcal{M}^*_{\alpha'\beta\gamma'\delta}
  .
\end{align}
where $\mathcal{M}_{\alpha\beta\gamma\delta}$ is the amplitude for the hard partonic process, see Fig.~\ref{fig:matel}.

For an unpolarised final hadron, the off-diagonal elements of $\mathcal{D}_{h/c}^{\gamma\gamma'}$ vanish, \emph{i.e.}, $\mathcal{D}_{h/c}^{\gamma\gamma'}\propto\delta_{\gamma\gamma'}$. Helicity conservation then implies $\alpha=\alpha'$, so there is no dependence on the spin of hadron~$A$ and all SSA's are zero. To avoid such a conclusion, either intrinsic quark transverse motion, or higher-twist effects must be invoked.

\subsection{Intrinsic transverse motion}
Quark intrinsic transverse motion can generate SSA's in three essentially different ways (a necessarily $T$-odd effect):
\begin{enumerate}
\item\label{it:Sivers}
for $\Vec{k}_T$ in hadron $A$, $f_a(x_a)$ is replaced by $\mathcal{P}_a(x_a,\Vec{k}_T)$, which may depend on the spin of $A$ (distribution level);
\item\label{it:Collins}
$\Vec\kappa_T$ in hadron $h$ allows $\mathcal{D}_{h/c}^{\gamma\gamma'}$ to be non-diagonal (fragmentation level);
\item\label{it:Boer}
for $\Vec{k}'_T$ in hadron $B$, $f_b(x_b)$ is replaced by $\mathcal{P}_b(x_b,\Vec{k}'_T)$. The spin of $b$ in the unpolarised $B$ may couple to the spin of $a$ (distribution level).
\end{enumerate}
The three corresponding mechanisms are: (\ref{it:Sivers}) the \citeauthor{Sivers:1989cc} effect~\cite{Sivers:1989cc}; (\ref{it:Collins}) the \citeauthor{Collins:1993kk} effect~\cite{Collins:1993kk}; (\ref{it:Boer}) an effect studied by \citeauthor{Boer:1999mm}~\cite{Boer:1999mm} in Drell-Yan processes. Note that all such intrinsic-$\Vec{k}_T$, -$\Vec\kappa_T$ or -$\Vec{k}'_T$ effects are $T$-odd; \emph{i.e.}, they require ISI or FSI. Note too that when transverse parton motion is included, the QCD factorisation theorem is not completely proven, but see~\cite{Ji:2004wu}.

Assuming factorisation to be valid, the cross-section is
\begin{align}
  E_h \, \frac{\D^3\sigma}{\D^3\Vec{P}_h} &=
  \sum_{abc} \sum_{\alpha\alpha'\beta\beta'\gamma\gamma'}
  \int \! \D{x}_a \,
  \D{x}_b \,
  \D^2\Vec{k}_T \,
  \D^2\Vec{k}'_T \,
  \frac{\D^2\Vec\kappa_T}{\pi z}
  \nonumber
\\
  & \qquad \null \times
  \mathcal{P}_a(x_a, \Vec{k}_T) \, \rho^a_{\alpha'\alpha} \,
  \mathcal{P}_b(x_b, \Vec{k}'_T) \, \rho^b_{\beta'\beta}
  \left(
    \frac{\D\hat\sigma}{\D\hat{t}}
  \right)_{\alpha\alpha'\beta\beta'\gamma\gamma'}
  \mathcal{D}_{h/c}^{\gamma'\gamma}(z, \Vec\kappa_T)
  ,
\end{align}
where again
\begin{align}
  \left(
    \frac{\D\hat\sigma}{\D\hat{t}}
  \right)_{\alpha\alpha'\beta\beta'\gamma\gamma'}
  =
  \frac1{16\pi\hat{s}^2} \, \sum_{\beta\delta}
  \mathcal{M}_{\alpha\beta\gamma\delta} \,
  \mathcal{M}^*_{\alpha'\beta\gamma'\delta}
\end{align}
The Sivers effect relies on $T$-odd $k_T$-dependent distribution functions and predicts an SSA of the form
\begin{align}
  &
  E_h \, \frac{\D^3\sigma( \Vec{S}_T)}{\D^3\Vec{P}_h} -
  E_h \, \frac{\D^3\sigma(-\Vec{S}_T)}{\D^3\Vec{P}_h}
  \nonumber
\\
  & \qquad \null =
  | \Vec{S}_T | \,
  \sum_{abc}
  \int \! \D{x}_a \,
  \D{x}_b \,
  \frac{\D^2\Vec{k}_T}{\pi z}
  \Delta_0^T{f}_a(x_a, \Vec{k}_T^2) \,
  f_b(x_b) \,
  \frac{\D\hat\sigma(x_a, x_b, \Vec{k}_T)}{\D\hat{t}}\,
  D_{h/c}(z)
  ,
\end{align}
where $\Delta_0^T{f}$ (related to $f_{1T}^\perp$) is a $T$-odd distribution.

\subsection{Higher twist}
\citet*{Efremov:1984ip} showed that non-vanishing SSA's can also be obtained in QCD by resorting to higher twist and the so-called gluonic poles in diagrams involving $qqg$ correlators. Such asymmetries were later evaluated in the context of factorisation by \citeauthor*{Qiu:1991pp.x}, who studied direct photon production \cite{Qiu:1991pp.x} and hadron production \cite{Qiu:1998ia}.
This program has been extended by \citet*{Kanazawa:2000hz.x} to the chirally-odd contributions. The various possibilities are:
\begin{align}
  \D\sigma &=
  \sum_{abc}
  \left\{ \vphantom{D_{h/c}^{(3)}}
    G_F^a(x_a, y_a) \otimes f_b(x_b) \otimes
    \D\hat\sigma \otimes D_{h/c}(z)
  \right.
  \nonumber
\\
  & \hspace{6em} \null +
  \DT{f}_a(x_a) \otimes E_F^b(x_b, y_b) \otimes
  \D\hat\sigma' \otimes D_{h/c}(z)
  \nonumber
\\[1ex]
  & \hspace{10em} \null +
  \left.
    \DT{f}_a(x_a) \otimes f_b(x_b) \otimes
    \D\hat\sigma'' \otimes D_{h/c}^{(3)}(z)
  \right\}
\end{align}
The first term is the chirally-even three-parton correlator pole mechanism, as proposed in \cite{Efremov:1984ip} and studied in \cite{Qiu:1991pp.x,Qiu:1998ia}; the second contains transversity and is the chirally-odd contribution analysed in \cite{Kanazawa:2000hz.x}; and the third also contains transversity but requires a \mbox{twist-3} fragmentation function $D_{h/c}^{(3)}$.

\subsection{Phenomenology}
\citet{Anselmino:2002pd} have compared data with models inspired by the previous possible ($k_T$-dependent) mechanisms and find good descriptions although they cannot differentiate between mechanisms. The calculations by \citet*{Qiu:1991pp.x} (based on three-parton correlators) also compare well but are complex. However, the \mbox{twist-3} correlators (as in $g_2$) obey constraining relations with $k_T$-dependent densities and also exhibit a novel factorisation property.

\subsection{Pole factorisation}
\begin{figure}[hbt]
  \centering
  \includegraphics[height=25mm,bb=142 661 250 749,clip]{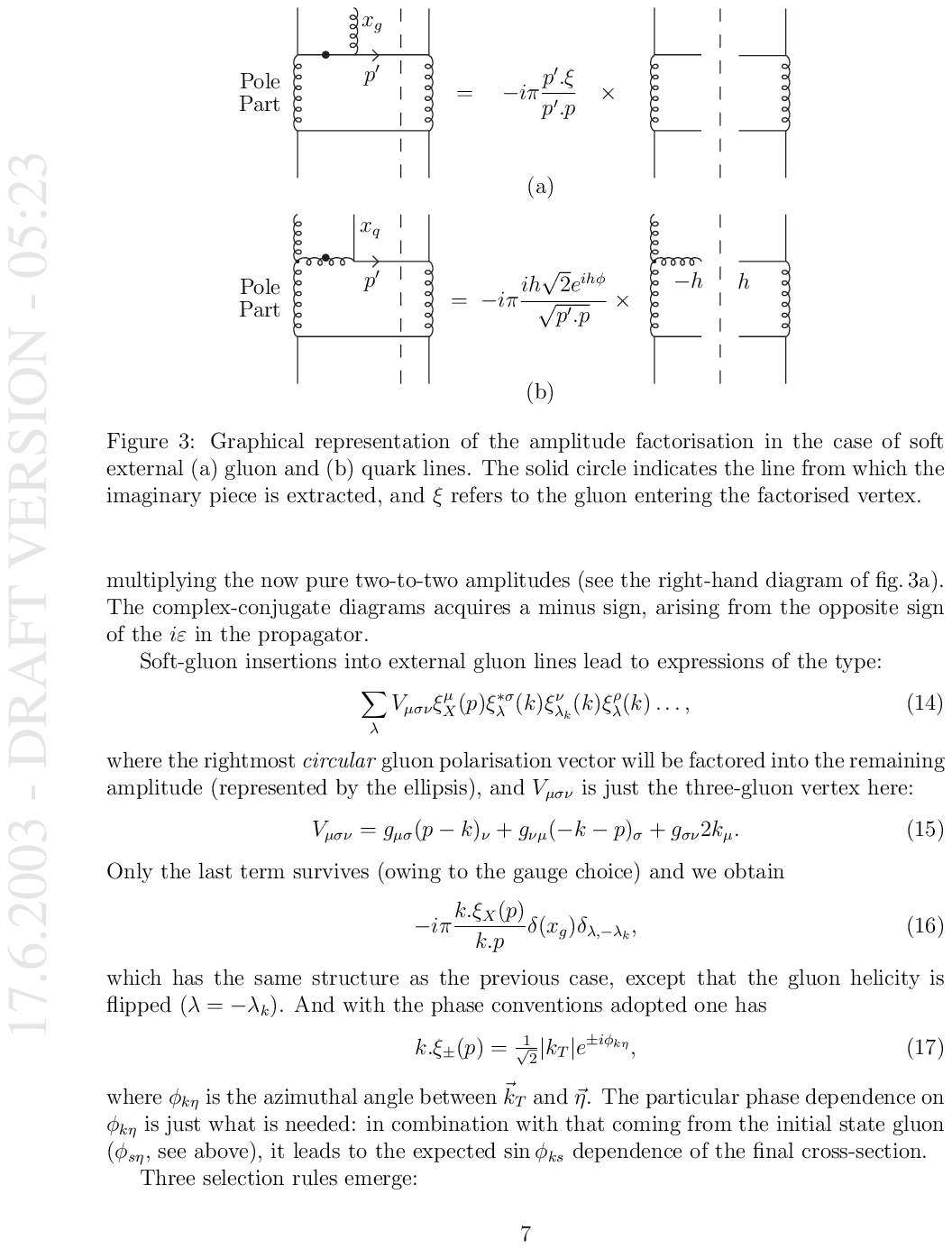}
  \includegraphics[height=25mm,bb=250 661 437 749,clip]{epsfiles/pole-factor}
  \caption{%
    An example of Pole factorisation: $p$ is the incoming proton momentum, $p'$ the outgoing hadron and $\xi$ is the gluon polarisation vector (lying in the transverse plane).
  }
  \label{fig:polefact}
\end{figure}

\citet*{Efremov:1984ip} noticed that the \mbox{twist-3} diagrams involving three-parton
\begin{wrapfigure}[14]{l}{30mm}
  \vspace*{-5mm}
  \centering
  \includegraphics[width=30mm,bb=92 558 196 701,clip]
                  {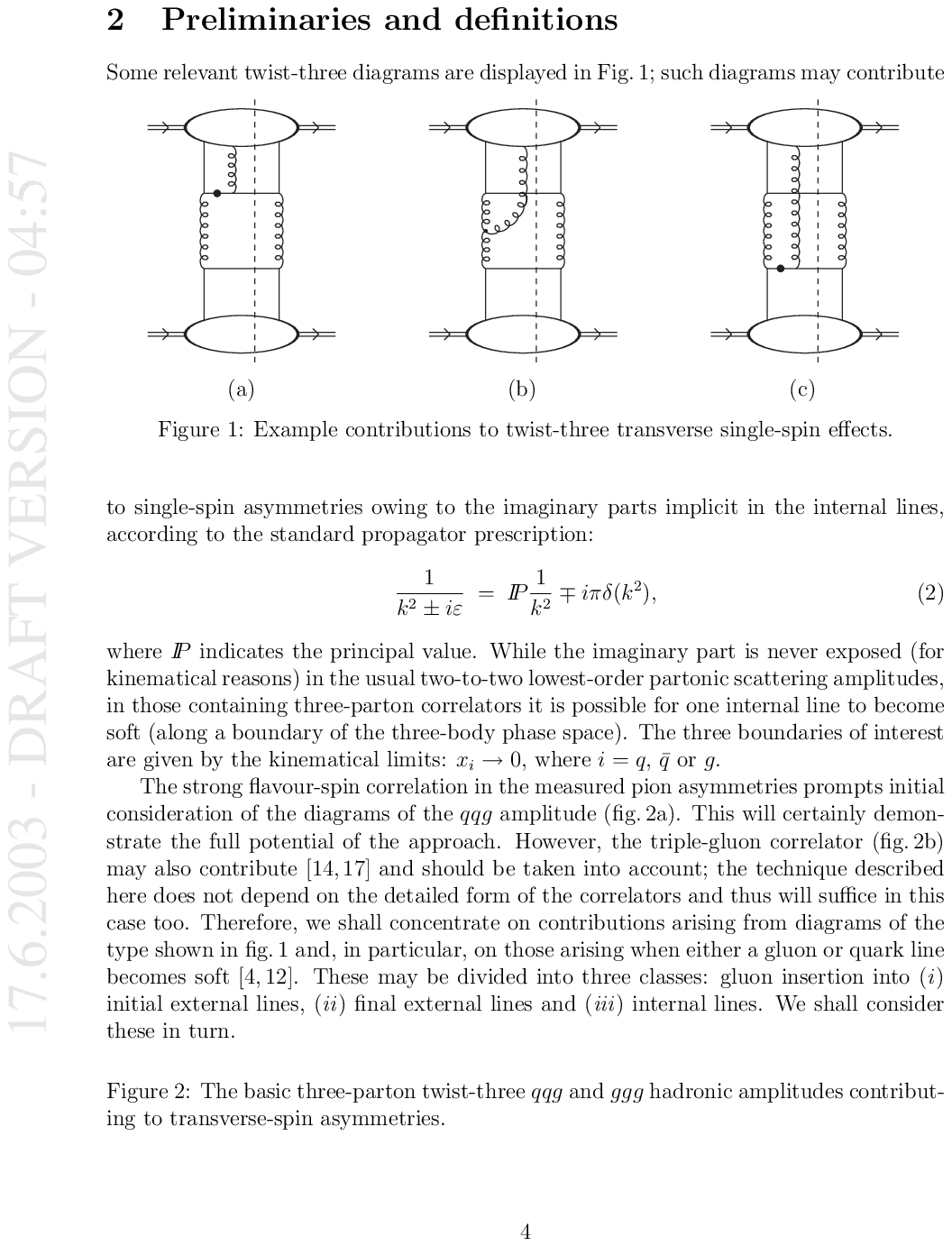}
  \caption{%
    Example of a propagator pole in a three-parton diagram.
  }
  \label{fig:propagator-pole}
\end{wrapfigure}
correlators can supply the necessary imaginary part via a pole term; spin-flip is implicit (and due to the gluon). The standard propagator prescription ($-\mkern-7mu\bullet\mkern-7mu-$ in Fig.~\ref{fig:propagator-pole}),
\begin{align}
  \frac{1}{k^2\pm\im\varepsilon} =
  \Principal \frac1{k^2} \mp \im\pi\delta(k^2)
\end{align}
leads to an imaginary contribution for $k^2\to0$.
A gluon with $x_gp$ inserted into an (initial or final) external line $p'$ sets $k=p'-x_gp$ and thus $x_g\to0$ $\Leftrightarrow$ $k^2\to0$, see Fig.~\ref{fig:polefact}.

This can be performed systematically for all poles (gluon and fermion): on all external legs with all insertions \citep{Ratcliffe:1998pq}.
The structures are still complex: for a given correlator there are many insertions, leading to different signs and momentum dependence.

The colour structure of the various diagrams (with the types of different soft insertions) is also different (we shall examine this question shortly). In all cases (examined) it turns out that just one diagram dominates in the large-$\Nc$ \begin{wrapfigure}[13]{r}{30mm}
  \vspace*{-5mm}
  \centering
  \includegraphics
    [width=30mm,bb=391 558 495 701,clip]
    {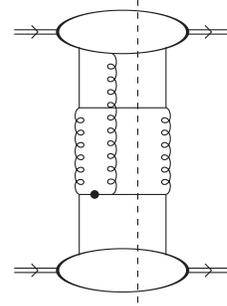}
  \caption{%
    Example of a dominant propagator pole diagram.
  }
  \label{fig:dominant}
\end{wrapfigure}
limit, see Fig.~\ref{fig:dominant}.
All other insertions into external (on-shell) legs are relatively suppressed by $1/\Nc^2$. This has been examined in detail by Ramilli (Insubria Univ.\ Masters thesis\nocite{Ramilli:2007t1}): the leading diagrams provide a good approximation. It has yet to be repeated for all the other \mbox{twist-3} contributions (\emph{e.g.}, also in fragmentation).

A question now arises: what is the relationship between the \mbox{twist-3} and $k_T$-dependent mechanisms? It might be hoped that, via the equations of motion \emph{etc.}, unique predictions for single-spin azimuthal asymmetries could be obtained by linking the (\emph{e.g.}, Sivers- or Collins-like) $k_T$-dependent mechanisms to the (Efremov--Teryaev) higher-twist three-parton mechanisms. \citet{Ma:2003ut} made a first attempt for the Drell-Yan process, but the predictions were found not to be unique. \citet{Ji:2006ub.x} have also examined the relationships between $k_T$-dependent and higher-twist mechanisms by matching in the intermediate $k_T$ region of common validity.

\section{More on Multiparton Correlators}

\subsection{Colour modification}
In \cite{Ratcliffe:2007ye} we provided an \emph{a posteriori} proof of the relation between \mbox{twist-3} and the Sivers function.
The starting point is a factorised formula for the Sivers function:
\begin{align}
  d\Delta\sigma
  &\sim
  \int \! d^2k_T^{} dx \,
  f_{\text{S}}^{} (x,k_T^{}) \,
  \epsilon^{\rho s P k_T^{}}
  \Tr\!\left[\vphantom{\big|} \gamma_\rho \, H(xP,k_T^{}) \right]
  .
\\
\intertext{%
  Expanding the subprocess coefficient function $H$ in powers of $k_T^{}$ and keeping the first non-vanishing term leads to
}
  &\sim
  \int \!\! d^2k_T^{} dx \,
  f_{\text{S}}^{} (x,k_T^{}) \,
  k_T^\alpha \, \epsilon^{\rho s P k_T^{}}
  \Tr\!\left[
    \gamma_\rho \, \frac{\partial H(xP,k_T^{})}{\partial k_T^\alpha}
  \right]_{k_T^{}{=}0}
  \mkern-35mu
  .
\end{align}
Exploiting various identities and the fact that there are other momenta involved, this can be rearranged into the following form:
\begin{equation}
  d\Delta\sigma \sim
  M \int \! dx \, f_{\text{S}}^{(1)}(x) \,
  \epsilon^{\alpha s P n}
  \Tr\!\left[
    \slashed{P} \, \frac{\partial H(xP,k_T^{})}{\partial k_T^\alpha}
  \right]_{k_T^{}{=}0}
  \mkern-35mu
  ,
\end{equation}
where
\begin{equation}
  f_{\text{S}}^{(1)}(x)
  = \int \! d^2k_T^{} \, f_{\text{S}}^{}(x,k_T^{}) \, \frac{k_T^2}{2M^2}
  .
\end{equation}
This coincides with the master formula of \cite{Koike:2006qv.x} for \mbox{twist-3} gluonic poles in high-$p_T^{}$ processes. The Sivers function can thus be identified with the gluonic-pole strength $T(x,x)$ multiplied by a process-dependent colour factor.

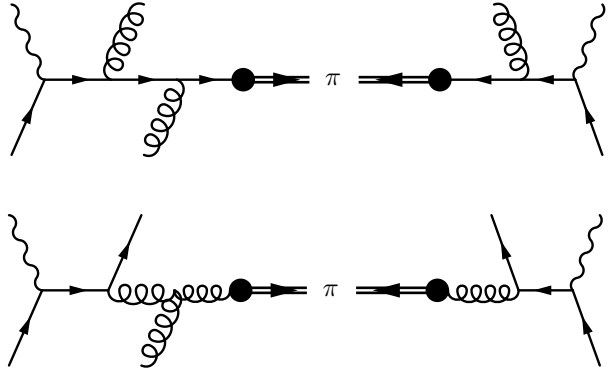
\begin{figure}[hbt]
  \centering
  \begin{fmffile}{fmffiles/SiversCol}%
    \fmfset{arrow_len}{2.5mm}%
    \fmfset{curly_len}{2.2mm}%
    \begin{fmfgraph*}(44,20)
      \fmfleft{i1,i2}
      \fmfright{d1,d2,d3}
      \fmftop{o2}
      \fmfbottom{o1}
      \fmf{phantom}{i1,v1,v2,o1}
      \fmf{phantom}{i2,v1}
      \fmf{phantom}{v2,o2}
      \fmffreeze
      \fmf{fermion}{i1,v1,v2,v3,v4}
      \fmf{dbl_plain_arrow}{v4,d2}
      \fmffreeze
      \fmf{photon}{v1,i2}
      \fmf{gluon}{o2,v2}
      \fmf{gluon}{o1,v3}
      \fmfv{d.sh=circle,d.fi=full,d.si=4thick}{v4}
      \fmfv{l=$\pi$}{d2}
    \end{fmfgraph*}
    \hspace{1.2em}
    \begin{fmfgraph}(36,20)
      \fmfright{i1,i2}
      \fmfleft{d1,d2,d3}
      \fmftop{o2}
      \fmfbottom{o1}
      \fmf{phantom}{i1,v1,v2,o1}
      \fmf{phantom}{i2,v1}
      \fmf{phantom}{v2,o2}
      \fmffreeze
      \fmf{fermion}{i1,v1,v2,v4}
      \fmf{dbl_plain_arrow}{v4,d2}
      \fmffreeze
      \fmf{photon}{v1,i2}
      \fmf{gluon}{v2,o2}
      \fmfv{d.sh=circle,d.fi=full,d.si=4thick}{v4}
    \end{fmfgraph}
    \\[5ex]
    \hspace*{0.15em}
    \begin{fmfgraph*}(44,20)
      \fmfleft{i1,i2}
      \fmfright{d1,d2,d3}
      \fmftop{o2}
      \fmfbottom{o1}
      \fmf{phantom}{i1,v1,v2,o1}
      \fmf{phantom}{i2,v1}
      \fmf{phantom}{v2,o2}
      \fmffreeze
      \fmf{fermion}{i1,v1,v2}
      \fmf{gluon}{v2,v3,v4}
      \fmf{dbl_plain_arrow}{v4,d2}
      \fmffreeze
      \fmf{photon}{v1,i2}
      \fmf{fermion}{v2,o2}
      \fmf{gluon}{o1,v3}
      \fmfv{d.sh=circle,d.fi=full,d.si=4thick}{v4}
      \fmfv{l=$\pi$}{d2}
    \end{fmfgraph*}
    \hspace{1.2em}
    \begin{fmfgraph}(36,20)
      \fmfright{i1,i2}
      \fmfleft{d1,d2,d3}
      \fmftop{o2}
      \fmfbottom{o1}
      \fmf{phantom}{i1,v1,v2,o1}
      \fmf{phantom}{i2,v1}
      \fmf{phantom}{v2,o2}
      \fmffreeze
      \fmf{fermion}{i1,v1,v2}
      \fmf{gluon}{v4,v2}
      \fmf{dbl_plain_arrow}{v4,d2}
      \fmffreeze
      \fmf{photon}{v1,i2}
      \fmf{fermion}{v2,o2}
      \fmfv{d.sh=circle,d.fi=full,d.si=4thick}{v4}
    \end{fmfgraph}
  \end{fmffile}
  \caption{%
    \mbox{Twist-3} SIDIS $\pi$ production via quark and gluon fragmentation.
  }
  \label{fig:SIDIS-3}
\end{figure}

The sign of the Sivers function depends on which of ISI or FSI is relevant:
\begin{equation}
  f_{\text{S}}^{(1)}(x) = \sum_i C_i \, \frac{1}{2M} T(x,x),
\end{equation}
where $C_i$ is a relative colour factor defined with respect to an Abelian subprocess. Now, to generate high $p_T^{}$, emission of an extra hard gluon is necessary. Then, according to the process under consideration, the FSI may occur \emph{before} or \emph{after} this emission, again leading to different colour factors. In this sense, factorisation is broken in SIDIS, although in a simple and accountable manner. In Fig.~\ref{fig:SIDIS-3} we consider the application of this relation to high-$p_T^{}$ SIDIS.

\subsection{Asymptotic behaviour}
The relation between gluonic poles, \emph{e.g.}, the Sivers function, and T-even transverse-spin effects, \emph{e.g.}, $g_2$ \citep{Shuryak:1981pi, Efremov:1983eb, Bukhvostov:1984as, Ratcliffe:1985mp, Balitsky:1987bk}, remains unclear. There are model-based estimates and approximate sum rules. However, the compatibility of general \mbox{twist-3} evolution \nocite{Bukhvostov:1984as, Ratcliffe:1985mp, Balitsky:1987bk} with dedicated studies of gluonic-pole evolution (\citet{Kang:2008ey,Zhou:2008mz} and at NLO \citet{Vogelsang:2009pj}) is still unproven. In the large-$x$ limit the evolution equations for $g_2$ diagonalise in the double-moment arguments \citep{Ali:1991em}. For the Sivers function and gluonic poles, this is the important kinematical region~\citep{Qiu:1991pp.x}.

The gluonic-pole strength $T(x)$, corresponds to a specific matrix element \citep{Qiu:1991pp.x}. It is also the residue of a general $qqg$ vector correlator $b_V(x_1,x_2)$~\cite{Korotkiian:1995vf}:
\begin{equation}
  b_V(x_1,x_2) =
  \frac{T(\frac{x_1+x_2}{2})}{x_1-x_2} + \text{regular part}
  ,
\end{equation}
which is defined as
\begin{equation}
  b_V(x_1,x_2) =
  \frac{i}{M} \! \int \! \frac{d\lambda_1 d\lambda_2}{2\pi} \,
  e^{i\lambda_1(x_1-x_2)+i\lambda_2 x_2} \,
  \epsilon^{\mu s p_1 n} \!
  \braket{p_1,s|
    \bar\psi(0) \, \slashed{n} \, D_\mu(\lambda_1) \, \psi(\lambda_2)
  |p_1,s}
  .
\end{equation}
There is another correlator, projected onto an axial Dirac matrix:
\begin{equation}
  b_A(x_1,x_2) =
  \frac1{M} \! \int \! \frac{d\lambda_1 d\lambda_2}{\pi}
  e^{i\lambda_1(x_1-x_2)+i\lambda_2x_2}
  \braket{p_1,s|
    \bar\psi(0) \, \slashed{n} \, \gamma^5 \,
    s{\cdot}D(\lambda_1) \, \psi(\lambda_2)
  |p_1,s}
  .
\end{equation}
This is required to complete the description of transverse-spin effects, in both SSA's and $g_2$. The two correlators have opposite symmetry properties for $x_1\leftrightarrow{}x_2$:
\begin{equation}
  b_A(x_1,x_2) =  b_A(x_2,x_1), \qquad
  b_V(x_1,x_2) = -b_V(x_2,x_1)
  ,
\end{equation}
which are determined by $T$ invariance.

In both DIS and in SSA's only a particular combination appears~\citep{Efremov:1983eb}:
\begin{equation}
  b_-(x_1,x_2) = b_A(x_2,x_1) - b_V(x_1,x_2)
  .
\end{equation}
The QCD evolution equations \citep{Bukhvostov:1984as, Ratcliffe:1985mp, Balitsky:1987bk} are usually written in terms of another quantity, which is expressed as matrix elements of the gluon field strength:
\begin{equation}
  Y(x_1,x_2) = (x_1-x_2) \, b_-(x_1,x_2)
  .
\end{equation}
It should be safe to assume that $b_-(x_1,x_2)$ has no double pole and
thus
\begin{equation}
  T(x) = Y(x,x)
  .
\end{equation}

Evolution is easiest studied in Mellin-moment form; for $Y(x,y)$ these become double moments:
\begin{equation}
  Y^{mn} = \int dx \, dy \, x^m \, y^n \, Y(x,y)
  ,
\end{equation}
where the variables are restricted to $|x|$, $|y|$ and $|x-y|<1$. We wish to examine the behaviour for $x$ and $y$ both close to unity and therefore close to each other. Thus, the gluonic pole provides the dominant contribution:
\begin{equation}
  \lim_{x,y\to1} Y(x,y) = T(\tfrac{x+y}{2}) + O(x-y)
  .
\end{equation}
In this approximation (now large $m,n$) the LO evolution equations simplify:
\begin{equation}
  \frac{d}{ds} Y^{nn} = 4\left(\CF+\frac{\CA}{2}\right)
  \ln{n} \, Y^{nn}
  ,
\end{equation}
where the evolution variable is $s=\beta_0^{-1}\ln\ln{Q^2}$.
In terms of $T(x)$ this is
\begin{equation}
  \dot T(x) =
  4\left(\CF+\frac{\CA}{2}\right)
  \int_x^1 dz \,
  \frac{(1-z)}{(1-x)}
  \frac1{\,(z-x)_+\!} \, T(z)
  ,
\end{equation}
which is similar to the unpolarised case, but differs by a colour factor $\left(\CF+\sfrac{\CA}{2}\right)$ and a softening factor $(1-z)/(1-x)$.

The extra piece in the colour factor ($\CA/2$) with respect to the unpolarised case ($\CF$) reflects the presence of a third active parton---the gluon. That is, the pole structure of three-parton kernels is identical, but the effective colour charge of the extra gluon is $\CA/2$. The softening factor is inessential to the asymptotic solution, it simply implies standard evolution for the function $f(x)=(1-x)T(x)$.
For an initial distribution $f(x,Q_0^2)=(1-x)^a$, the asymptotic solution \citep{Gross:1974fm} is the same but modified by $a\to{}a(s)$, with
\begin{equation}
  a(s) = a + 4\left(\CF+\frac{\CA}{2}\right) s
  .
\end{equation}
For $T(x)$, $a$ shifts to $a-1$; the evolution modification is identical; the spin-averaged asymptotic solutions are thus also valid for $T(x)$. This large-$x$ limit of the evolution agrees, bar the colour factor itself, with studies of gluonic-pole evolution~\citep{Kang:2008ey, Zhou:2008mz, Vogelsang:2009pj}.

\section{Summary and Conclusions}

Viewing the Sivers function as an effective \mbox{twist-3} gluonic-pole contribution \cite{Ratcliffe:2007ye}, it is seen to be process dependent: besides a sign (ISI \emph{vs.} FSI), there is a process-dependent colour factor. This factor is determined by the colour charge of the initial and final partons. It generates the sign difference between SIDIS and Drell-Yan at low $p_T^{}$, but in hadronic reactions at high $p_T^{}$ it is more complicated. Such a picture is complementary to the matching in the region of common validity. Such matching between various $p_T^{}$ regions now takes the form of a $p_T^{}$-dependent colour factor. It also lends some justification to the possibility of global Sivers function fits~\citep{Teryaev:2005bp}.

We have also shown the applicability of generic \mbox{twist-3} evolution equations to the Sivers function. Its effective nature allows us to relate the evolution of T-odd (Sivers function) and T-even (gluonic pole) quantities. An important ingredient of our approach is the large-$x$ approximation, where gluonic-poles dominate and the evolution simplifies. We have found that the Sivers function evolution is multiplicative and described by a colour-factor modified \mbox{twist-2} spin-averaged kernel~\cite{Ratcliffe:2009r1}.

\section*{Acknowledgments}

In closing, let me (belatedly) wish Anatoly a very happy 75th.\ birthday and also thank the organisers for this excellent workshop. We should also thank Gregory Korchemsky and Feng Yuan for useful discussions. The funding for the visits of O.T. to Como is due in part to the Landau Network (Como) and also to the recently concluded PRIN2006 on Transversity. Some of the ideas discussed have already been presented at other workshops~\citep*{Ratcliffe:2007p1, Ratcliffe:2008p2}.


\end{document}